\pdfoutput=1
\documentclass[aps,prb,twocolumn,10pt,showpacs,amsmath,amssymb,floatfix]{revtex4-1}
\usepackage{amsmath}
\usepackage{graphicx}
\usepackage{verbatim}
\usepackage{color}
\usepackage{subfigure}
\usepackage{hyperref}
\usepackage{soul}
\usepackage{amssymb}
\usepackage{cancel}
\usepackage{float}
\begin{document}
\title{Generic ordering of structural transitions in quasi-one-dimensional Wigner crystals}

\author{J. E. \surname{Galv\'an-Moya}} \email[Email: ]{JesusEduardo.GalvanMoya@uantwerpen.be}
\affiliation{Department of Physics, University of Antwerp, Groenenborgerlaan 171, B-2020, Antwerp, Belgium}

\author{V. R. \surname{Misko}}
\affiliation{Department of Physics, University of Antwerp, Groenenborgerlaan 171, B-2020, Antwerp, Belgium}

\author{F. M. \surname{Peeters}} \email[Email: ]{Francois.Peeters@uantwerpen.be}
\affiliation{Department of Physics, University of Antwerp, Groenenborgerlaan 171, B-2020, Antwerp, Belgium}

%
\begin{abstract}
We investigate the dependence of the structural phase transitions in an infinite
 quasi-one-dimensional system of repulsively interacting particles on the profile of the confining
 channel.
 Three different functional expressions for the confinement potential related to real experimental
 systems are used that can be tuned continuously from a parabolic to a hard-wall potential in
 order to find a thorough understanding of the ordering of the chain-like structure transitions.
 We resolve the longstanding issue why the most theories predicted a 1-2-4-3-4 sequence of chain
 configurations with increasing density, while some experiments found the 1-2-3-4 sequence.
\end{abstract}
\maketitle

\section{Introduction}
A crystalline structure consists of a periodic arrangement of molecules, atoms or different
 particles.  The first prediction about self-arrangement of particles, nowadays known as Wigner
 crystals (WC)\cite{045_wigner}, states that in the absence of kinetic energy, a system of
 interacting particles arranges itself into a body-centered cubic (bcc) lattice in
 three-dimensional (3D) space\cite{008_hasse,005_cornelissens}, a triangular lattice in
 two-dimensions (2D)\cite{004_schweigert,009_partoens,040_meyer}, while in
 one-dimensions (1D), the energetically most favorable organization is given by an
 evenly spaced lattice\cite{127_schulz,128_deshpande,040_meyer}.

For a quasi-one-dimensional (Q1D) system, Piacente \emph{et al.}\cite{165_piacente,003_piacente}
 studied the ground state configuration (GS) of a system of particles confined in a parabolic
 channel, and found a non-sequential ordering of transitions (non-SOT) between
 $1-2-4-3-4-5-6$-chain-like structures with increasing particle density.
 They revealed that this ordering of transitions between chains is robust, being not affected by
 the range of the interaction between the particles\cite{173_piacente,006_piacente,041_galvan}.
 That succession of phases differs from a sequential ordering of transitions (SOT), which is
 characterized by a consecutive succession of phases with
 $1-2-3-4-5-6$-chains, as one would intuitively expect to be the case.
The structural transition from two- to four-chain configuration occurs, in case of a non-SOT,
 through a zigzag transition
 of each of the two chains and a simultaneous small shift along the chain, which
 makes it a discontinuous transition\cite{003_piacente}.

The only second order transition in this sequence is the zigzag transition between one-
 and two-chain configuration, which has been extensively studied in
 classical\cite{014_fishman,006_piacente,011_delcampo,041_galvan,168_ikegami} and
 quantum\cite{038_meyer,040_meyer,039_meng,166_shimshoni,167_bermudez,170_cormick} systems.
 A detailed analysis of the structural transitions for larger number of chains has to a lesser
 extent also been addressed\cite{165_piacente,003_piacente,147_koppl,171_klironomos}. 
 Experimental findings in a colloidal Q1D system showed evidence of transitions from eight- up to
 five-chain configurations~\cite{147_koppl}. Numerical calculations in the same
 work, suggested that this sequence continues reducing the number of chains, one by one, until
 the three-chain configuration. No information was provided about the transition between two- and
 three-chain structures.

However, a direct transition from two- to three-chain configuration, $2-3$, has been shown to take
 place in a number of systems. For example, for Yukawa particles the direct $2-3$-transition was
 observed in dusty plasma clusters~\cite{037_sheridan} with increasing linear density.  
 In addition, a SOT has been predicted theoretically for an Abrikosov-vortex arrangement in a
 superconducting slab for low
 temperatures~\cite{193_guimpel,194_brongersma,195_carneiro,192_sardella,172_barba}, for Pearl
 vortices~\cite{134_bronson}, and also for binary mixtures of repulsive
 particles~\cite{012_ferreira}, in particular, when the ratio between charges of both species was
 around $1/5$. 

These examples suggest that, in spite of the demonstrated robustness of the non-SOT with respect
 to the range of the interaction between the particles, the non-sequential ordering is perhaps
 sensitive to the system parameters and conditions.  First, the real confinement can be different
 from parabolic. For example, in case of colloids~\cite{147_koppl} the boundaries could be closer
 to hard walls.
 In case of superconducting vortices, the potential barrier preventing vortices from entering or
 escaping the slab is described by the known Bean-Livingston barrier~\cite{B09_Tinkham,187_bean},
 which for a wide slab is very different from a parabola. 
 In addition, fluctuations of any nature can be responsible for the disappearance of the non-SOT.
 This was probably the case in the experiment~\cite{169_ikegami} that analyzed the melting of the
 WC chain-like structures and their transport in a Q1D channel of electrons on a liquid He surface.
 In particular, the non-SOT has been observed in that experiment for very low temperatures, while
 even at T=1K the non-SOT regime was washed out, and the usual SOT was observed instead. 
 This behavior is in agreement with the early predictions by Piacente \emph{et al.}~\cite{003_piacente}
 showing that thermal fluctuations can easily destroy the non-SOT. 
 This finding is also in agreement with recent computer simulations~\cite{188_vasylenko} on the
 dynamics of WC in Q1D channels with constrictions. It was shown that even in the absence of
 thermal fluctuations, the non-SOT observed in long constrictions was destroyed in short
 constrictions, due to fluctuations of the number of particles flowing through the constriction.
 
Although fluctuations are generally a universal ``tool'' to destroy any ordering, and as shown in
 the examples above also the non-SOT, the role of other factors such as the functional form of the
 confinement potential remains unexplored. 
In particular, an important open question is: How universal is the non-SOT? Is it typical for
 systems with parabolic confinement, or is it of a more generic nature? The positive answer to the
 latter would open broader possibilities for experimental observation of the non-SOT, provided the
 fluctuations are very weak. 
This motivated us to investigate the universality of the non-SOT. 

In the present work we study the influence of the confinement potential on the GS of a Q1D system
 of interacting particles, elucidating the general model of the order of the transitions between
 chain-like structures.  Different confinement potentials are used in order to study the behavior
 of the GS transitions, when the profile of the channel is varied continuously from a
 parabolic-like to a hard-wall potential.

\section{Model System}

We consider an infinite system of identical interacting particles with mass $m$ and charge $q$,
 which are trapped in a Q1D channel through an external confinement potential, restricting the
 movement of the particles in the $y$ direction. 
 The total energy of the system is given by the following expression: 
\begin{equation}
 H = \sum_{i=1}^{\infty} \sum_{j>i}^{\infty}
	    V_{int}(|\mathbf r_{ij}|) + \sum_{i=1}^{\infty} V_{conf}(y_{i}),
\end{equation}
 where $\mathbf r_{ij}=\mathbf r_{i} - \mathbf r_{j}$ is the relative position of the $i$-th with
 respect to the $j$-th particle in the system, while $V_{int}(r)$ and $V_{conf}(y)$ represent the
 pairwise inter-particle interaction and the confinement potential of the channel, respectively. 
 
The inter-particle interaction is taken as follows:
\begin{eqnarray} \label{interaction_types}
  V_{int}(r) & = & \frac{q^2}{\epsilon R} \frac{R^n e^{-r/\lambda}}{r^n}, 
\end{eqnarray}
 where the parameters $\lambda$ and $n$ allow us to tune the range of the interaction between
 particles in the system, while $\epsilon$ is the dielectric constant of the medium the particles
 are moving in and $R$ is a parameter with dimension of length. 
In order to understand the effect of confinement on the ordering of the phase transitions, we
 considered the following three different functional forms for the confinement potential:
\begin{eqnarray}
  V_{A}(\alpha,y) & = & \frac{m\upsilon_A^2 y_{0}^{2}}{2}
			  \left|\frac{y}{y_0}\right|^{\alpha}, \label{conf_alpha_orig} \\
  V_{B}(\beta,y)  & = & \frac{m\upsilon_A^2 y_{0}^{2}}{2}
			  \frac{\cosh(\beta y) - 1}{\cosh(\beta y_0)-1}, \label{conf_cosh_orig} \\
  V_{C}(\gamma,y) & = & \frac{m\upsilon_C^2 y_{0}^{2}}{2}
			  \left[ \text{e}^{-\gamma^2(y-y_{0})^2}
			       + \text{e}^{-\gamma^2(y+y_{0})^2} \label{conf_gauss_orig} \right],
\end{eqnarray}
 where $y_0$ determines the effective width of the confinement channel, while the parameters
 $\alpha$, $\beta$ and $\gamma$ allow to control the sharpness of its profile.
\begin{figure}[htpb!]
\begin{center}
\subfigure{\includegraphics[trim={0.0cm 0 0 0}, scale=0.70]{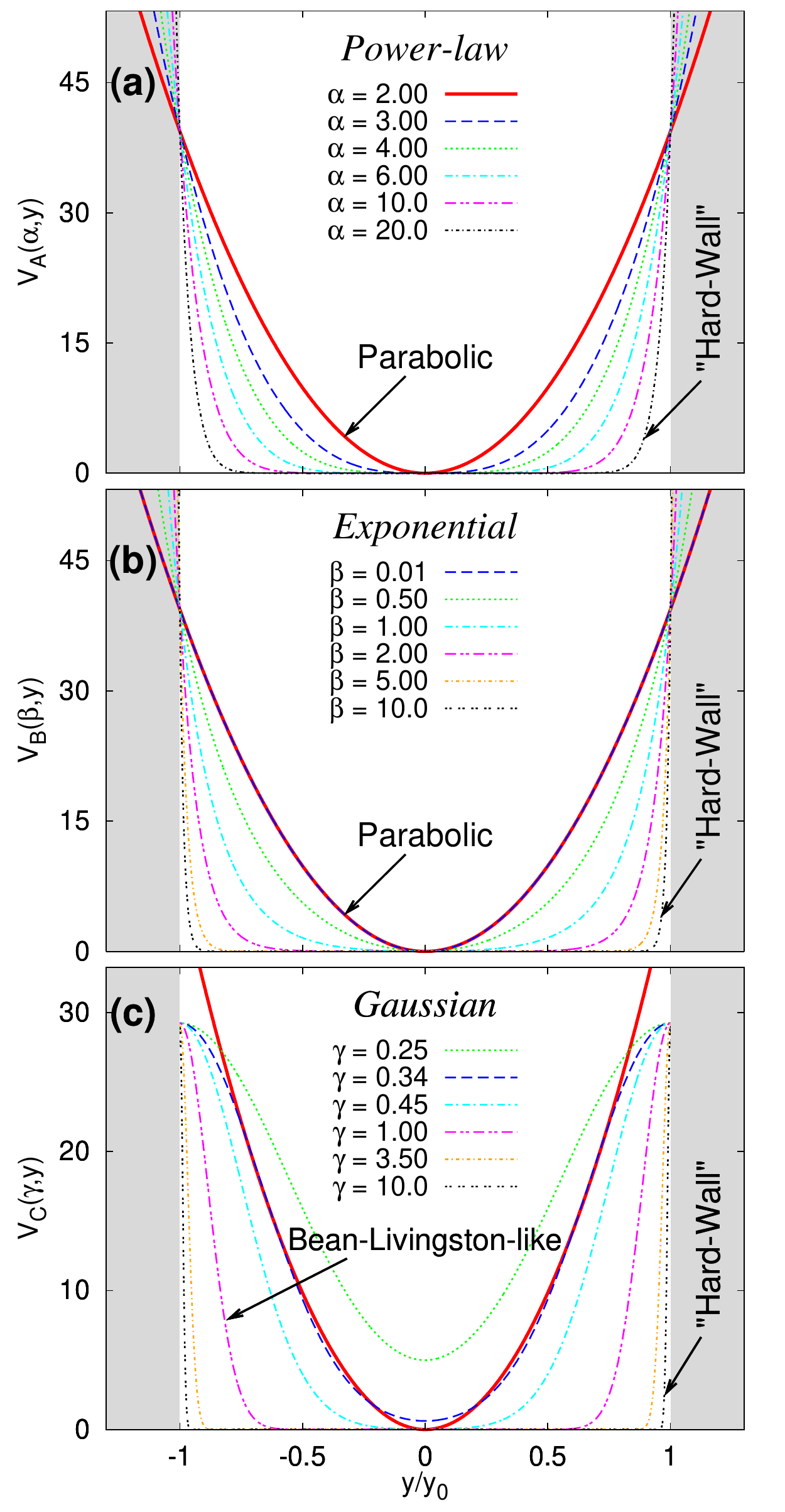}}
\caption{\label{fig:vpots} (Color online) Profile of the confinement potentials considered in this
 work, for different values of the shape parameters: a) power-law, b) exponential-like, c) Gaussian
 (see details in the text).
 The solid red curve shows the parabolic potential for reference, and the hard-wall potential is indicated
 by the white region in the middle of the figure.}
\end{center}
\end{figure}

In dimensionless form, the interaction and the confinement potentials in our model become: 
\begin{eqnarray}
  V_{int}(r) & = & \frac{e^{\kappa r}}{r^n}, \\
  V_{A}(\alpha,y) & = & \upsilon^2 y_{0}^{2-\alpha} \left|y\right|^{\alpha}, \label{conf_alpha}\\
  V_{B}(\beta,y)  & = & \upsilon^2 y_{0}^{2}
			    \frac{\cosh(\beta y) - 1}{\cosh(\beta y_0)-1}, \label{conf_cosh} \\
  V_{C}(\gamma,y) & = & \sigma^2 y_{0}^{2}
			      \left[ \text{e}^{-\gamma^2(y-y_{0})^2}
				   + \text{e}^{-\gamma^2(y+y_{0})^2} \label{conf_gauss} \right],
\end{eqnarray}
where the energy is expressed in units of
 $E_0 = (m\omega_0^2/2)^{n/(n+2)} (q^2/\epsilon)^{2/(n+2)} R^{2(n-1)/(n+2)}$ and all distances
 are expressed in units of $r_0 = (2q^2/m\omega_0^2\epsilon)^{1/(n+2)} R^{(n-1)/(n+2)}$. 
 The dimensionless frequencies are given by $\upsilon=\upsilon_A/\omega_0$ and
 $\sigma=\upsilon_C/\omega_0$, while $\omega_0$ measures the strength of the confinement
 potential, and the screening of the pairwise interaction is $\kappa=r_0/\lambda$.  The
 dimensionless linear density $\eta$ is defined as the number of particles per unit of length
 along the unconfined direction.
\begin{figure}[htpb!]
\begin{center}
\subfigure{\includegraphics[trim={0.0cm 0 0 0}, scale=0.60]{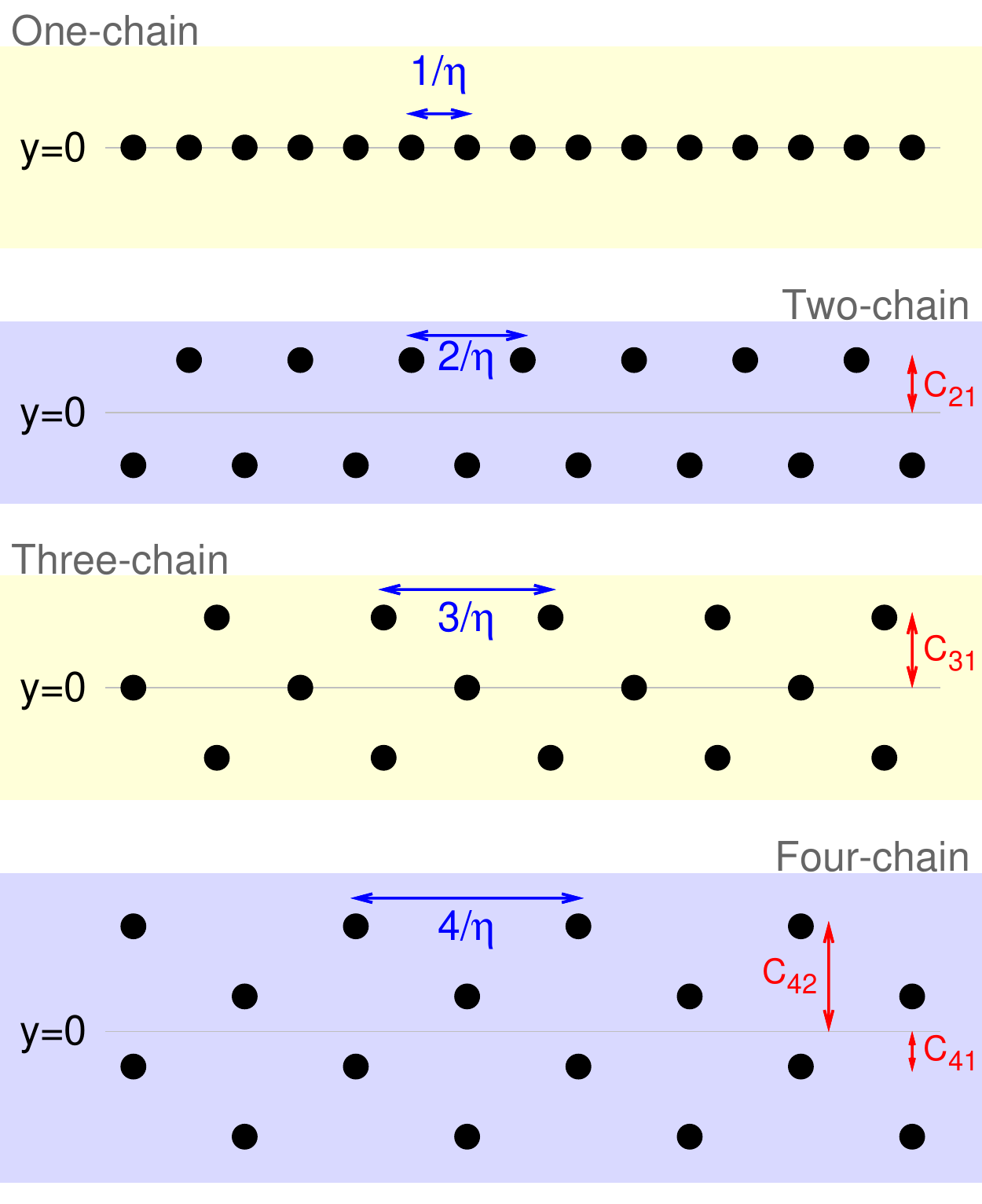}}
\caption{\label{fig:conf_chains} (Color online) Schematic view of the chain-like configurations
 together with the relevant order parameters used in the calculations, where $\eta$ represents
 the linear particle density in the unconfined direction.}
\end{center}
\end{figure}
 
Previously, it has been shown that the sequence of transitions are not affected by the range of
 the interaction between particles\cite{003_piacente}. Therefore, in the present work the range of
 the interaction is fixed by choosing $\kappa=1$ and $n=1$. 
In order to analyze the influence of the profile of the channel on the sequence of the phase
 transitions, we fix the confinement strength for the parabolic potential (i.e. $\alpha=2$ in
 Eq.~(\ref{conf_alpha})) to $\upsilon=1$.
Next we note that $V_{B}(\beta=0,y)=V_{A}(\alpha=2,y)$ is a parabola.  For $V_{C}(\gamma,y)$ we
 determine the parameters $\sigma$ and $y_0$ such that for some $\gamma$-value we obtain a
 confinement potential that is very close to a parabola (see Fig.~\ref{fig:vpots}(c)).
The fitting results in the following choice of parameters $\sigma=0.862$ and $y_0=6.275$, which
 are fixed for all numerical calculations performed in the present work.
The different confinement potentials are plotted in Fig.~\ref{fig:vpots} for different values of
 the shape parameters.
 
Fig.~\ref{fig:vpots} shows the flexibility of the confinement potentials defined by
 Eqs.~(\ref{conf_alpha}-\ref{conf_gauss}). Thus, by gradually changing the shape parameters
 $\alpha$, $\beta$, and $\gamma$, we follow a continuous evolution from a soft parabolic-like 
 (note that in Fig.~\ref{fig:vpots}(c), due to the shape of the confinement potential given by
 Eq.~(\ref{conf_gauss}), the closest approximation to the parabolic confinement is found for
 $\gamma=0.34$, this profile is being plotted with a blue dashed line) to the hard-wall potential.
 As a reference, the parabolic profile is shown by the red curve in each plot.
It is worth noting that the three functional forms for the confinement potential are essentially
 different and approach the hard-wall limit in a different manner. 
The potential profiles were chosen such that they model confinement potentials in various
 physical systems ranging from charged particles and colloids in narrow channels to vortices in
 superconducting stripes.
For example, the potential profile in Fig.~\ref{fig:vpots}(c) for $\gamma=1$ models the
 Bean-Livingston barrier for vortex exit from a superconductor.

\section{Transitions between chain-like structures}
The GS of the system of interacting particles in a Q1D channel consists of chain-like
 structures~\cite{014_fishman,006_piacente,011_delcampo,041_galvan}, and the transitions between
 them are of first order~\cite{003_piacente,147_koppl,171_klironomos}, with the exception of the
 zigzag transition between 1 to 2 chains which is of second order~\cite{006_piacente,041_galvan}.
 Some typical chain-like configurations are shown in Fig.~\ref{fig:conf_chains}, where the order
 parameters are indicated in red. 
 
In order to study the structural transitions in an infinite system of particles interacting
 through the potential $V_{int}(r)$ and confined in a Q1D channel defined by $V_{conf}(y)$, we
 calculate the total energy of the system for some typical $n$-chains structures, as follows: 
\begin{widetext}
\begin{eqnarray}
  H_{1ch} & = &	  V_{conf}\left(0\right)
	        + \sum_{m=1}^{\infty} V_{int}\left(\frac{m}{\eta} \right), \\
  H_{2ch} & = &    	      V_{conf}\left(\frac{c_{21}}{\eta} \right)
	        + \sum_{m=1}^{\infty} V_{int}\left(\frac{2m}{\eta} \right)
	        + \sum_{m=1}^{\infty} V_{int}\left(\frac{2}{\eta}
				  \sqrt{c_{21}^2+\left(m-\frac{1}{2}\right)^2} \right), \\
  H_{3ch} & = &   \frac{1}{3}V_{conf}\left(0 \right)
	        + \frac{2}{3}V_{conf}\left( \frac{3c_{31}}{\eta} \right)
	        + \frac{1}{3}V_{int}\left(\frac{6c_{31}}{\eta} \right)
	        +            \sum_{m=1}^{\infty} V_{int}\left(\frac{3m}{\eta} \right)
	        + \frac{2}{3}\sum_{m=1}^{\infty} V_{int}\left(\frac{3}{\eta}
				  \sqrt{4c_{31}^2+m^2} \right)  \nonumber \\ & &
		+ \frac{4}{3}\sum_{m=1}^{\infty} V_{int}\left(\frac{3}{\eta}
				  \sqrt{c_{31}^2+\left(m-\frac{1}{2}\right)^2} \right),
				  \>\>\>\>\>\>\>\>\>\>\>\\
  H_{4ch} & = &   \frac{1}{2}V_{conf}\left(\frac{4c_{41}}{\eta} \right)
	        + \frac{1}{2}V_{conf}\left(\frac{4c_{42}}{\eta} \right)
	        + \frac{1}{2}V_{int}\left(\frac{4\left(c_{41}+c_{42}\right)}{\eta}\right)
	        +            \sum_{m=1}^{\infty} V_{int}\left(\frac{4m}{\eta} \right)  \nonumber \\ & &
	        +            \sum_{m=1}^{\infty} V_{int}\left(\frac{4}{\eta}
				  \sqrt{ \left(c_{41}+c_{42}\right)^2 + m^2}\right)
	        + \frac{1}{2}\sum_{m=1}^{\infty} V_{int}\left(\frac{4}{\eta}
				  \sqrt{ 4c_{41}^2 
				       + \left(m-\frac{1}{2}\right)^2}\right)  \nonumber \\ & &
	        + \frac{1}{2}\sum_{m=1}^{\infty} V_{int}\left(\frac{4}{\eta}
				  \sqrt{ 4c_{42}^2
				       + \left(m-\frac{1}{2}\right)^2}\right)  
	        +            \sum_{m=1}^{\infty} V_{int}\left(\frac{4}{\eta}
				  \sqrt{ \left(c_{42}-c_{41}\right)^2
				       + \left(m-\frac{1}{2}\right)^2}\right),
\end{eqnarray}
\end{widetext}
where all the distances are expressed in units of the distance between adjacent particles in each
 chain, as indicated in Fig.~\ref{fig:conf_chains}.

We find the GS of the system by minimizing the energy (numerically) with respect to the order
 parameter(s) for each chain-like structure.

\subsection{Power-law confinement} 
 
\begin{figure} [b!]
\begin{center}
\subfigure{\includegraphics[trim={0.0cm 0 0 0}, scale=0.68]{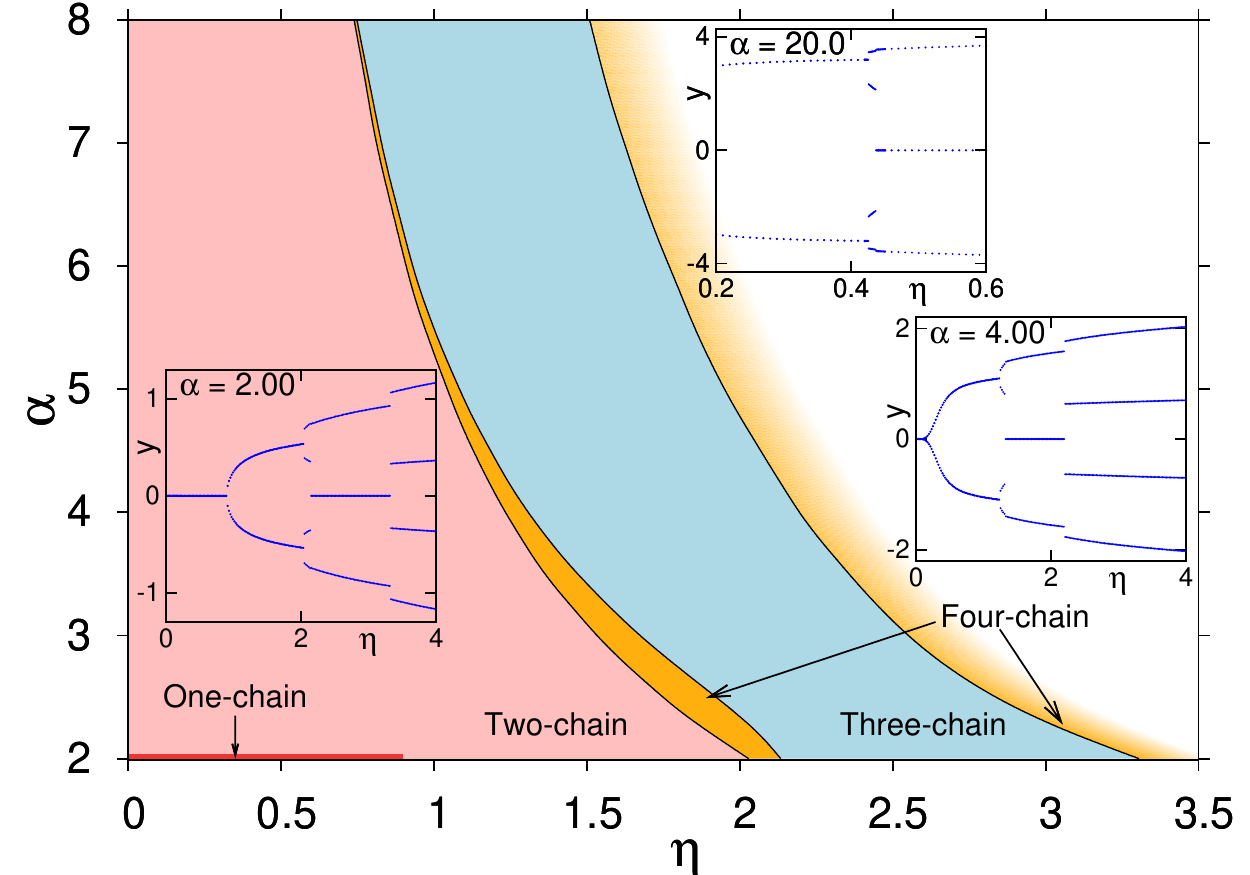}}
\caption{\label{fig:phdiag_alpha} (Color online) Phase diagram of the ground state for a system
 with confinement $V_{A}(\alpha,y)$, as function of $\alpha$ and the linear density $\eta$.  All
 phase transitions are of first order except the zigzag transition which is only possible for
 $\alpha=2$.  The insets show the $y$ position of the particles as function of $\eta$ for
 different values of $\alpha$ as indicated in each figure.}
\end{center}
\end{figure}

In the case of power-law confinement (Eq.~(\ref{conf_alpha})), the phase diagram of the GS is shown
 in Fig.~\ref{fig:phdiag_alpha} as a function of the shape parameter $\alpha$ and the linear
 density $\eta$.  As was analytically demonstrated in
 Refs.~[\onlinecite{006_piacente},\onlinecite{041_galvan}],
 the stability of the one-chain configuration as the GS is only guaranteed for the case of
 $\alpha=2$, while for larger values of $\alpha$ the one-chain configuration is no longer found as
 the GS.  This result is represented in Fig.~\ref{fig:phdiag_alpha} by the thick red line (for
 $\alpha=2$) showing the small region ($0<\eta<0.9$) where the one-chain structure is found as the
 GS.  The $y$-position of the particles forming the GS, for $\alpha=2$, is presented as a function
 of $\eta$ in the left-hand side inset of Fig.~\ref{fig:phdiag_alpha}, where the non-SOT is
 clearly present (i.e., the transitions $2 - 4 - 3$). 
 
For $\alpha>2$ and small $\eta$, the two-chain configuration is the GS of the system even at low
 densities, where the inter-chain distance (i.e., the order parameter $c_{21}$) slowly decreases
 but never becomes exactly zero~\cite{041_galvan} except for $\eta\rightarrow 0$. 
 This behavior is illustrated in the right-hand side inset in Fig.~\ref{fig:phdiag_alpha} for
 $\alpha=4$, where one can also see that a direct transition between two- and three-chain
 configuration is not found (i.e., non-SOT).  Indeed, a small region where the four-chain
 arrangement is the GS remains between the two- and the three-chains structures, even for large
 values of the shape parameter $\alpha$.  Thus the GS transition between two-, four- and
 three-chain configurations is still present for $\alpha=20$ (i.e., close to the hard-wall limit), as shown
 in the upper inset in Fig.~\ref{fig:phdiag_alpha}.

With further increasing the density, the three-chain GS configuration is found as the ordered sequence
 of the GS configurations with the number of chains increasing one by one (the transitions between
 the states are of first order), i.e., three-, four-, five-, six-chain, etc. 

Therefore, a system of particles confined by a power-law potential (Eq.~(\ref{conf_alpha})), shows
 a robust non-sequential transition between two-, four-, and three-chain configurations as the
 ground states, for a broad range of densities $\eta$ and the shape of the confinement profile
 varying from parabolic to hard-wall.

\subsection{Exponential-type of confinement}

The parabolic potential is the simplest and often used form to model quasi-one dimensional
 systems of, e.g., charged particles~\cite{165_piacente,003_piacente}, colloids~\cite{147_koppl}
 and dusty plasma~\cite{037_sheridan,191_tkachenko}. Another useful form of the confinement
 potential is the one with exponentially decaying barriers, as described by Eq.~(\ref{conf_cosh})
 that uses a hyperbolic cosine, where $\beta$ acts as a parameter which controls the shape of the
 channel.  The advantage of this form of confinement potential is that, by tuning the control
 parameter $\beta$, it rapidly evolves into a flat central part in the potential profile providing
 a fast continuous transition to a hard-wall-like profile (see Fig.~1(b)). 
  
\begin{figure} [t!]
\begin{center}
\subfigure{\includegraphics[trim={0.0cm 0 0 0}, scale=0.68]{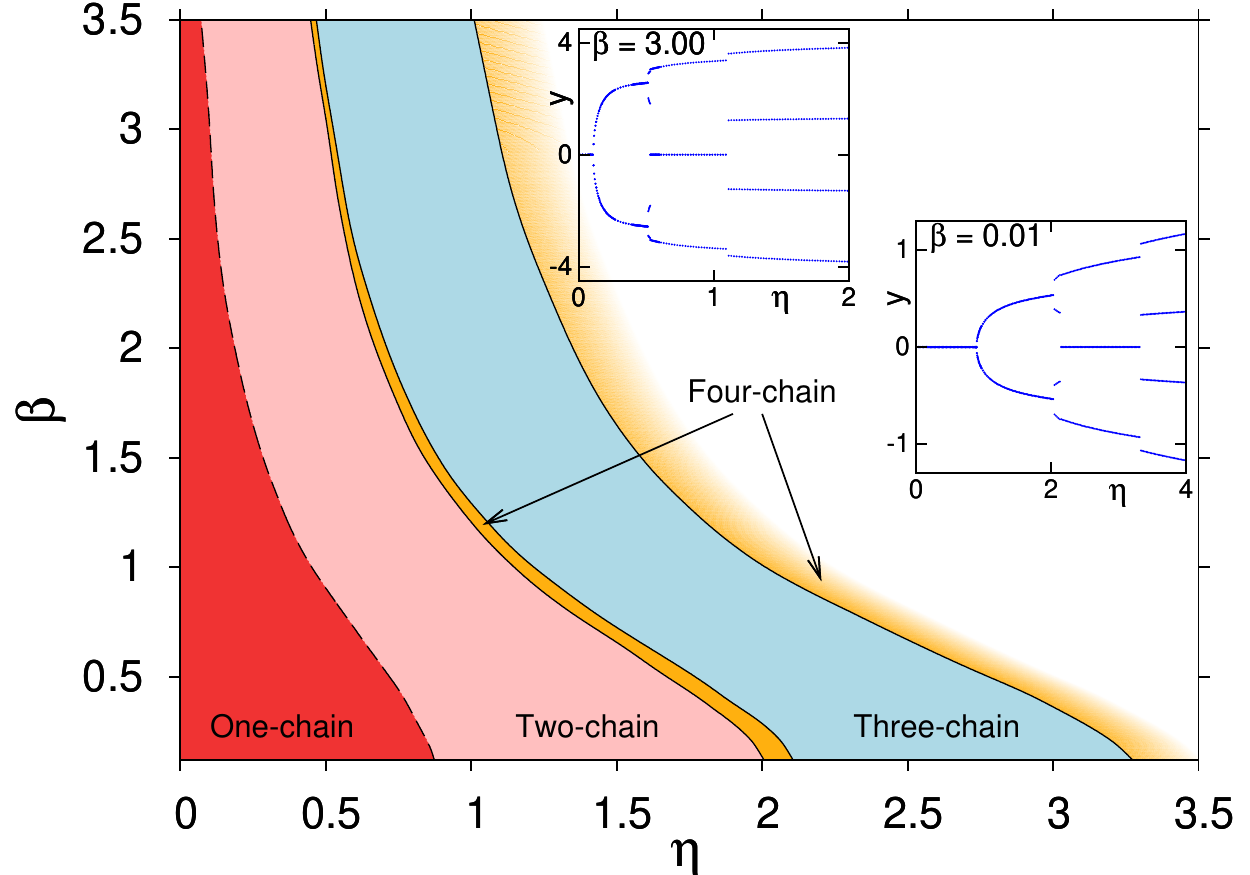}}
\caption{\label{fig:phdiag_cosh} (Color online) Phase diagram of the ground state for a system
 with confinement $V_{B}(\beta,y)$, as function of $\beta$ and the linear density $\eta$.  The
 solid and dashed lines represent first and second order transitions, respectively.  The insets
 show the $y$ position of the particles as function of $\eta$ for different values of $\beta$
 as indicated in each figure.}
\end{center}
\end{figure}

The phase diagram for the GS of a system of particles confined by the potential $V_B(\beta,y)$ is
 shown in Fig.~\ref{fig:phdiag_cosh} as a function of the parameters $\beta$ and $\eta$. 

The limiting case of a parabolic potential is recovered by setting $\beta=0$. 
The $y$-coordinate of the particles as a function of $\eta$ is shown in the right-hand side inset
 of Fig.~\ref{fig:phdiag_cosh} for $\beta=0.01$. 
With increasing $\beta$, the transitions between chain-like configurations occur at lower values
 of the density, and the transition between one- and two-chain configuration (zigzag transition)
 is of second order (indicated by the dashed curve in Fig.~\ref{fig:phdiag_cosh}).
Opposite to the above case of a parabolic confinement, the zigzag transition is always stable, even for large values of $\beta$. 

Therefore, the evolution of the GS of the system is guided by a non-SOT irrespective of the value
 of $\beta$, thus allowing the emergence of the four-chain state between the two- and three-chain
 configurations.  Although the width of this intermediate four-chain region slowly decreases with
 increasing $\beta$, it does not disappear when the channel profile approaches the hard-wall
 potential, as shown for $\beta=3$ at the upper inset in Fig.~\ref{fig:phdiag_cosh}. 
It is worth noting that, although this profile evolution (i.e., from a parabolic to a hard-wall)
 is qualitatively similar to that for the power-law confinement (compare Figs.~1(a) and (b)), the
 power-law confinement does not show the state with one chain (except for $\alpha=2$ and narrow range of
 $\eta$), and the four-chain state (between three- and two-chains) rapidly shrinks with increasing
 $\alpha$.
The latter can be the reason that the non-SOT can hardly be detected in channels with power-law
 confinement in the limit of hard walls, and instead a usual SOT is observed (e.g., in dusty
 plasma~\cite{037_sheridan}).

\subsection{Gaussian confinement} 

An even softer transition between parabolic and hard-wall potential is presented in this section. 
 Here we present a model where the confinement is presented by two gaussians, symmetrically
 positioned with respect to the center of the channel.  The shape of the confinement  is controlled
 by the parameter $\gamma$, as shown in Eq.~(\ref{conf_gauss}).  For a specific value of
 $\gamma\approx 1$ it represents an approximation to the Bean-Livingston
 barrier~\cite{B09_Tinkham,187_bean} for vortices interacting with the boundary of a
 superconductor. 

\begin{figure} 
\begin{center}
\subfigure{\includegraphics[trim={0.0cm 0 0 0}, scale=0.68]{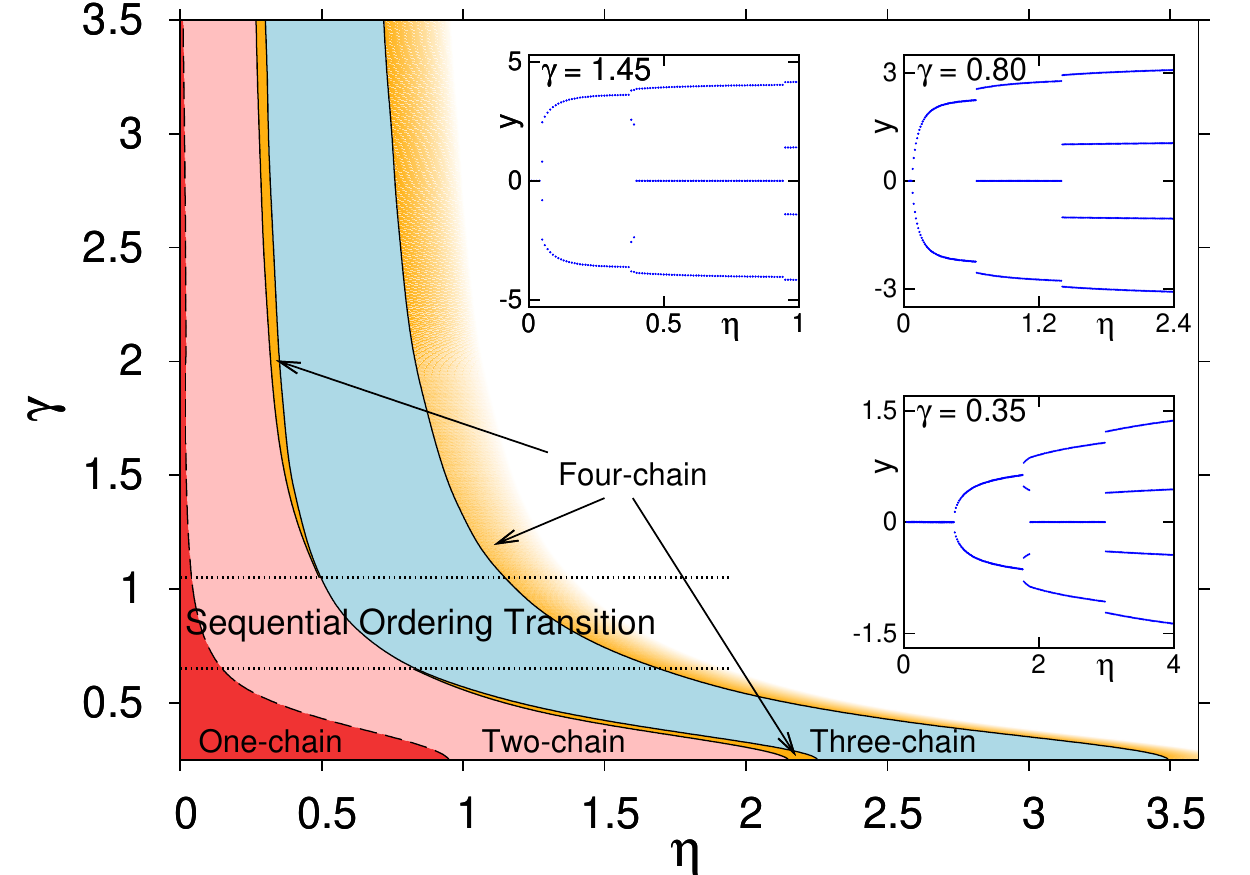}}
\caption{\label{fig:phdiag_gauss} (Color online) Phase diagram of the ground state for a system
 with confinement $V_{C}(\gamma,y)$, as function of $\gamma$ and the linear density $\eta$.  The
 solid and dashed lines represent first and second order transitions, respectively, while the
 region enclosed by the pointed lines indicates the sequential ordering transition.  The insets
 show the $y$ position of the particles as function of $\eta$ for different values of $\gamma$
 as indicated in each figure.}
\end{center}
\end{figure}

The phase diagram of a system of particles confined by $V_C(\gamma,y)$ is presented in
 Fig.~\ref{fig:phdiag_gauss} as a function of $\gamma$ and $\eta$.  The best fit to a parabolic
 potential is provided by choosing $\gamma=0.34$.  Then the evolution of the GS is guided by a
 non-SOT when density increases, as shown in the lower inset of Fig.~\ref{fig:phdiag_gauss} where
 the $y$-coordinate of the particles is plotted as a function of $\eta$. 
As one can expect, in this case the GS undergoes a similar series of transitions as in case of
 parabolic confinement.

However, as one can see from Fig.~\ref{fig:phdiag_gauss}, the narrow region of the intermediate
 four-chain structure gradually shrinks with increasing $\gamma$ and around $\gamma=0.61$ this
 region disappears thus allowing a straightforward first order transition from two- to three-chain
 configuration.   Therefore, we observe a SOT in the system, as shown in the upper right-hand side inset of
 Fig.~\ref{fig:phdiag_gauss} where the transversal position of the particles is plotted as a
 function of the linear density. 

On the other hand, this SOT is only observed as the GS for a certain range of the parameter
 $\gamma$, namely, from $\gamma=0.61$ to $\gamma=1.05$. For larger $\gamma$-values, the
 intermediate four-chain configuration is restored thus preventing the direct transition between
 two- and three-chain configurations. 
An important result is that this range includes the shape that fits the Bean-Livingston barrier for
 superconducting vortices. While we are not aware of any other explanations why the non-SOT has
 never been observed in superconducting slabs, our numerical result clearly indicates that non-SOT
 {\it does not exist} in a system of
 vortices confined by Bean-Livingston barriers in a superconducting stripe.

\section{Conclusions}

In this work, we studied the GS transitions of a system of particles interacting through a screened
 Yukawa potential and confined in a Q1D channel, where the structures found correspond to Wigner
 crystal configurations. 

The effect of the confinement on the GS transitions, for increasing system density, is analyzed
 for different confinement profiles: the power-law ($\sim|y|^\alpha$), exponential and Gaussian
 potentials modulating the transversal profile of the channel through a shape parameter. 
 Analytical expression for the energy of different $n$-chains configurations are calculated, and
 the GS is found by minimization of the energy with respect to the order parameter(s) of each
 analyzed structure. 

As reference limiting cases, we defined a parabolic (``soft'' confinement) and a hard-wall
 confinement, and the proposed potential profiles are able to transit continuously between these
 two limits.  While asymptotically resembling each other, the different profiles evolve in a
 different manner for intermediate values of the shape parameters.  This resulted in different
 sets of GS configurations which were analyzed in detail and summarized in phase diagrams ``shape
 parameter versus density'', for each considered confinement potential. 
Our choice of the model confinements was guided by those found in different physical systems, i.e.
 particles in a quasi-one dimensional channel, when increasing
 the channel width.  In particular, these correspond to charged particles in parabolic trap which
 are realized, e.g., in experiments with dusty plasmas, colloids confined in narrow channels, or
 even vortices in superconducting stripes. 

As follows from our analysis, due to the above similarity of the profiles for the limiting cases,
 all the systems display a similar behavior in the two limiting cases.  Thus, the ground states of the
 systems with parabolic-like confinement profiles always evolve following a non-SOT.
Similarly, all three systems allow a non-SOT in the hard-wall limit, although in the case of
 power-law confinement, as mentioned above, the one-chain configuration is missing. 
Simultaneously, we found that the non-SOT is present in all the systems for intermediate values of
 the shape parameter thus indicating that the non-SOT is extremely robust for a broad range of
 possible profiles and shape parameters.

At the same time, for the Gaussian confinement potential, a striking SOT for the GS was found to
 appear within a window of the shape parameter (i.e., $0.61<\gamma<1.05$).  It is worth noting that
 this window includes the shape that describes the Bean-Livingston barrier for superconducting
 vortices.  This result shows that the non-SOT does not exist in a system of vortices confined by
 Bean-Livingston barriers in a superconducting stripe and thus explains why the non-SOT, which was
 shown to be robust for different confinements, was never found for superconducting vortices. 

Note that in many physical systems under real conditions, fluctuations may destroy the intermediate
 four-chain configuration which is probably the reason why the SOT (but not the non-SOT) has been
 observed in several experiments with colloids, dusty plasmas, and electrons in narrow channels. 
 One indication that these thermal fluctuations, are responsible for destroying the non-SOT for
 the Wigner crystal, was found numerically where it was shown that the non-SOT is present only for
 very low temperatures. 
In addition, it was recently found that fluctuating number of particles in narrow short channels
destroys the non-SOT while it is present in long narrow channels. 

Thus our findings open the possibilities for using the confinement potential to manipulate the GS transition in the Q1D Wigner crystals and also can stimulate further studies in the field, both in theory and experiment.

\acknowledgments

This work was supported by the Flemish Science Foundation (FWO-Vl) and the Odysseus and
 Methusalem programmes of  the  Flemish  government.
Computational  resources were provided by HPC infrastructure of the University of Antwerp
 (CalcUA) a division of the Flemish Supercomputer Center (VSC).

%
%

\end{document}